\documentclass[12pt]{JHEP}

\def \B {B}

\def\uno{\mbox{1 \kern-.59em {\rm l}}}  

\newcounter{multieqs}

\newcommand{\bq}{\begin{equation}}
\newcommand{\fq}{\end{equation}}
\newcommand{\bqr}{\begin{eqnarray}}
\newcommand{\fqr}{\end{eqnarray}}

\newcommand{\be}{\begin{equation}}
\newcommand{\ee}{\end{equation}}
\newcommand{\eq}[1]{(\ref{#1})}

\newcommand{\ket}[1]{|#1 \rangle}

\newcommand{\bm}[1]{\mbox{\boldmath $#1$}}

\def\bd{\begin{document}}
\def\ed{\end{document}}
\def\nn{\nonumber}
\def\bea{\begin{eqnarray}}
\def\eea{\end{eqnarray}}
\let\bm=\bibitem
\let\la=\label
\renewcommand{\theequation}{\thesection.\arabic{equation}}
\def\npb#1#2#3{Nucl. Phys. {\bf{B#1}} #3 (#2)}
\def\plb#1#2#3{Phys. Lett. {\bf{#1B}} #3 (#2)}
\def\prl#1#2#3{Phys. Rev. Lett. {\bf{#1}} #3 (#2)}
\def\prd#1#2#3{Phys. Rev. {D \bf{#1}} #3 (#2)}
\def\cmp#1#2#3{Comm. Math. Phys. {\bf{#1}} #3 (#2)}
\def\cqg#1#2#3{Class. Quantum Grav. {\bf{#1}} #3 (#2)}
\def\nppsa#1#2#3{Nucl. Phys. B (Proc. Suppl.) {\bf{#1A}}#3 (#2)}
\def\ap#1#2#3{Ann. of Phys. {\bf{#1}} #3 (#2)}
\def\ijmp#1#2#3{Int. J. Mod. Phys. {\bf{A#1}} #3 (#2)}
\def\rmp#1#2#3{Rev. Mod. Phys. {\bf{#1}} #3 (#2)}
\def\mpla#1#2#3{Mod. Phys. Lett. {\bf A#1} #3 (#2)}
\def\jhep#1#2#3{J. High Energy Phys. {\bf #1} #3 (#2)}
\def\atmp#1#2#3{Adv. Theor. Math. Phys. {\bf #1} #3 (#2)}

\newcommand{\EQ}[1]{\begin{equation} #1 \end{equation}}
\newcommand{\AL}[1]{\begin{subequations}\begin{align} #1 \end{align}\end{subequations}}
\newcommand{\SP}[1]{\begin{equation}\begin{split} #1 \end{split}\end{equation}}
\newcommand{\ALAT}[2]{\begin{subequations}\begin{alignat}{#1} #2 \end{alignat}\end{subequations}}
\def\beqa{\begin{eqnarray}}
\def\eeqa{\end{eqnarray}}
\def\beq{\begin{equation}}
\def\eeq{\end{equation}}

\def\N{{\cal N}}
\def\sst{\scriptscriptstyle}
\def\thetabar{\bar\theta}
\def\Tr{{\rm Tr}}
\def\one{\mbox{1 \kern-.59em {\rm l}}}

\def\a{\alpha}          \def\da{{\dot\alpha}}
\def\b{\beta}           \def\db{{\dot\beta}}
\def\c{\gamma}  \def\C{\Gamma}  \def\cdt{\dot\gamma}
\def\d{\delta}  \def\D{\Delta}  \def\ddt{\dot\delta}
\def\e{\epsilon}                \def\vare{\varepsilon}
\def\f{\phi}    \def\F{\Phi}    \def\vvf{\f} \def\vphi{\varphi}
\def\h{\eta}
\def\k{\kappa}
\def\l{\lambda} \def\L{\Lambda}
\def\m{\mu}     \def\n{\nu}
\def\o{\omega}
\def\p{\pi}     \def\P{\Pi}
\def\r{\rho}
\def\s{\sigma}  \def\S{\Sigma}
\def\t{\tau}
\def\th{\theta} \def\Th{\Theta} \def\vth{\vartheta}
\def\X{\Xeta}
\def\z{\zeta}

\def\cA{{\cal A}} \def\cB{{\cal B}} \def\cC{{\cal C}}
\def\cD{{\cal D}} \def\cE{{\cal E}} \def\cF{{\cal F}}
\def\cG{{\cal G}} \def\cH{{\cal H}} \def\cI{{\cal I}}
\def\cJ{{\cal J}} \def\cK{{\cal K}} \def\cL{{\cal L}}
\def\cM{{\cal M}} \def\cN{{\cal N}} \def\cO{{\cal O}}
\def\cP{{\cal P}} \def\cQ{{\cal Q}} \def\cR{{\cal R}}
\def\cS{{\cal S}} \def\cT{{\cal T}} \def\cU{{\cal U}}
\def\cV{{\cal V}} \def\cW{{\cal W}} \def\cX{{\cal X}}
\def\cY{{\cal Y}} \def\cZ{{\cal Z}}

\def\ua{\underline{\alpha}}
\def\ub{\underline{\phantom{\alpha}}\!\!\!\beta}
\def\uc{\underline{\phantom{\alpha}}\!\!\!\gamma}
\def\um{\underline{\mu}}
\def\ud{\underline\delta}
\def\ue{\underline\epsilon}
\def\una{\underline a}\def\unA{\underline A}
\def\unb{\underline b}\def\unB{\underline B}
\def\unc{\underline c}\def\unC{\underline C}
\def\und{\underline d}\def\unD{\underline D}
\def\une{\underline e}\def\unE{\underline E}
\def\unf{\underline{\phantom{e}}\!\!\!\! f}\def\unF{\underline F}
\def\unm{\underline m}\def\unM{\underline M}
\def\unn{\underline n}\def\unN{\underline N}
\def\unp{\underline{\phantom{a}}\!\!\! p}\def\unP{\underline P}
\def\unq{\underline{\phantom{a}}\!\!\! q}
\def\unQ{\underline{\phantom{A}}\!\!\!\! Q}
\def\unH{\underline{H}}

\def\As {{A \hspace{-6.4pt} \slash}\;}
\def\bs {{b \hspace{-6.4pt} \slash}\;}
\def\Ds {{D \hspace{-6.4pt} \slash}\;}
\def\ds {{\del \hspace{-6.4pt} \slash}\;}
\def\ss {{\s \hspace{-6.4pt} \slash}\;}
\def\ks {{ k \hspace{-6.4pt} \slash}\;}
\def\ps {{p \hspace{-6.4pt} \slash}\;}
\def\pas {{{p_1} \hspace{-6.4pt} \slash}\;}
\def\pbs {{{p_2} \hspace{-6.4pt} \slash}\;}

\def\Fh{\hat{F}}
\def\Vh{\hat{V}}
\def\Xh{\hat{X}}
\def\ah{\hat{a}}
\def\xh{\hat{x}}
\def\yh{\hat{y}}
\def\ph{\hat{p}}
\def\xih{\hat{\xi}}

\def\psit{\tilde{\psi}}
\def\Psit{\tilde{\Psi}}
\def\tht{\tilde{\th}}

\def\At{\tilde{A}}
\def\Qt{\tilde{Q}}
\def\Rt{\tilde{R}}

\def\ct{\tilde{c}}
\def\ft{\tilde{f}}
\def\st{\tilde{s}}
\def\pt{\tilde{p}}
\def\qt{\tilde{q}}
\def\vt{\tilde{v}}

\def\delb{\bar{\partial}}
\def\bz{\bar{z}}
\def\bD{\bar{D}}
\def\bB{\overline{B}}

\def\bd{{\bf d}}
\def\bk{{\bf k}}
\def\bl{{\bf l}}
\def\bp{{\bf p}}
\def\bq{{\bf q}}
\def\br{{\bf r}}
\def\bx{{\bf x}}
\def\by{{\bf y}}
\def\bR{{\bf R}}
\def\bV{{\bf V}}

\def\d{\delta}\def\D{\Delta}\def\ddt{\dot\delta}

\def\pa{\partial} \def\del{\partial}
\def\xx{\times}

\def\trp{^{\top}}
\def\inv{^{-1}}
\def\dag{{^{\dagger}}}
\def\pr{^{\prime}}

\def\rar{\rightarrow}
\def\lar{\leftarrow}
\def\lrar{\leftrightarrow}

\newcommand{\0}{\,\!}      
\def\one{1\!\!1\,\,}
\def\im{\imath}
\def\jm{\jmath}

\newcommand{\tr}{\mbox{tr}}
\newcommand{\slsh}[1]{/ \!\!\!\! #1}

\def\vac{|0\rangle}
\def\lvac{\langle 0|}

\def\hlf{\frac{1}{2}}
\def\ove#1{\frac{1}{#1}}

\def\Box{\square}
\def\ZZ{\mathbb{Z}}
\def\CC#1{({\bf #1})}
\def\bcomment#1{}
\def\bfhat#1{{\bf \hat{#1}}}
\def\VEV#1{\left\langle #1\right\rangle}

\newcommand{\ex}[1]{{\rm e}^{#1}} \def\ii{{\rm i}}

\def\vs{\vspace*{.3cm}}   

\def\iin{{\rm n}}
\def\iim{{\rm m}}

\title{Noncommutative D-brane and Open String
in pp-wave Background with B-field}

\author{Chong-Sun Chu$^{a}$, Pei-Ming Ho$^{b}$\\
$^a$Centre for Particle Theory, Department of Mathematical Sciences,
University of Durham, Durham, DH1 3LE, UK\\
$^b$ Department of Physics,
National Taiwan University, Taipei 106, Taiwan, R.O.C. \\
E-mail: {\tt chong-sun.chu@durham.ac.uk, pmho@phys.ntu.edu.tw}
}

\abstract{
The open string ending on a D-brane with a constant $B$-field
in a pp-wave Ramond-Ramond background is exactly solvable.
The theory is controlled by three dimensionful parameters: $\a'$,
the mass parameter (RR background times the lightcone momentum) and
the $B$-field.  We quantize the open string 
theory and determine the full noncommutative structure.  
In particular, we find a fully noncommutative phase space whose
noncommutativity depends on all these parameters.
The lightcone Hamiltionian is obtained, and  as a consequence of  
the nontrivial commutation relations of the theory, new features of the spectrum
are noted.
Various scaling limits of the string results are considered. Physical
implications are discussed.   
}

\keywords{Non-Commutative Geometry, Open String, D-brane, pp-wave}

\preprint{{\tt hep-th/0203186}}    

\begin{document}  

\section{Introduction}

Recently a new maximal supersymmetric IIB supergravity background, the
so called pp-wave RR background, was
discovered \cite{bfhp1}. 
The pp-wave background consists of a plane wave metric, 
supported by a homogeneous RR 5-form flux
\bea 
&&ds^2 = -{\rm f}^2 x^i x^i  (dx^+)^2 + 2 dx^+ dx^- + dx^i dx^i, \quad
i = 1, \dots, 8, \label{pp1} \\
&&F_5 = {\rm f} dx^+ \wedge( dx^1 \wedge dx^2 \wedge dx^3 \wedge dx^4 + 
dx^5 \wedge dx^6 \wedge dx^7 \wedge dx^8 ). \label{pp2}
\eea
A constant Euclidean metric $g_{ij}$ can be
introduced easily. We will do so in section 3.
This background has 32 supersymmetries 
and is related to the $AdS_5\times S^5$ background \cite{bfhp1,bfhp2,bfhp3}
by a Penrose limit \cite{penrose}. 
Moreover it  is remarkable  that string theory in
this background is exactly solvable \cite{metsaev}.
The understanding of the properties of string
theory in this background is very valuable and is of great interest.
The Green-Schwarz formulation of closed string in pp-wave background has 
been performed by Metsaev \cite{metsaev} (see also \cite{t1,t2}). 
Boundary states for the lightcone GS strings in pp-wave
background has been constructed \cite{billo}. The covariant NSR
formulation remains illusive however.

It is  natural to ask whether there is a meaning of the  
``Penrose limit'' for the celebrated AdS/CFT correspondence \cite{malda}.
Based on the  spectrum of the lightcone Hamiltonian, 
Berenstein, Maldacena and Nastase   \cite{BMN} had put
forward a remarkable proposal that string theory on this pp-wave
background is dual to the large $N$ limit  of a certain subsector of
the 4-dimensional  supersymmetric $\cN=4$ $SU(N)$ gauge
theory. Various aspects as well as generalizations had been considered
in \cite{g1}--\cite{g8}.

D-branes are basic objects in  the nonperturbative formulation of string
theory. One of the developments in the last few years of string theory
is the realization that a constant $B$-field on a D-brane leads to
noncommutative geometry on the D-brane worldvolume
\cite{CDS,DH,CH1,SW}. In this setting, the $B$-field and the string
length square $\a'$ are both
dimensionful. Therefore, it is conceivable that a double scaling
limit may be taken so that certain
dimensionful parameter remains in the low energy limit  $\a' \to 0$. 
This is precisely what Seiberg and Witten did
\cite{SW} and remarkably they obtained as the low energy limit 
a noncommutative field theory with the dimensionful noncommutativity 
parameters $\th^{\m\n}$. It is an interesting question whether there
are other string backgrounds for which
new kind of noncommutative geometries arise.

Now string in pp-wave is exactly solvable (at least in the GS
formulation), and  there is a mass parameter $m$ 
(see \eq{mass} below) 
in addition to $\a'$.  If we also turn on
a $B$-field, then we  have {\it three} dimensionful parameters at our
disposal. It is therefore very natural
to consider a D-brane with a constant $B$-field sitting in the pp-wave
background and check whether any new kind of quantum geometry arises,
and to consider the possible scaling limits in the low energy.
This is the main motivation of our work. 

We recall that from the point of view of \cite{CH1,CH2}, 
noncommutative geometry on the D-brane worldvolume is a direct consequence of 
the open string mixed boundary condition that occurs due to the nonzero
$B$-field. This gives rise to a noncommutativity of the string zero
mode $x_0^i$ among themselves, but leaves the momentum zero mode
$p_0^i$ commuting  among themselves. As it turns out, we will find
that by turning on a mass parameter $m$, $p_0^i$ also becomes
noncommuting. Thus the phase space of the zero modes $x_0, p_0$
becomes fully noncommuting, see \eq{zCR1}--\eq{zCR3}. 
Moreover we find
that the noncommutativity depends on the lightcone momentum $p^+$. 
This is one of the main results of this paper. 
Its physical meaning is
intriguing and further understanding of it  will be important. 
An immediate consequence of the commutation relations we obtain 
is that the spectrum of the lightcone Hamiltonian is modified in a 
nontrivial manner by the $B$-field (see \eq{lcH2}). These are possible since $p^+$
is central in the pp-wave supersymmetry algebra.

The  paper is organized as follows. In section 2, we first present a
D-brane configuration that preserves half of the 32 supersymmetries. We
show that it is allowed to put a constant $B$-field on its
worldvolume. 
Section 3 is devoted to the quantization of the open string ending on
this D-brane.
In section 3.1, we construct the complete mode expansion of an open string
ending on a D-brane with $B$-field in the pp-wave background. The zero
mode part has  a nontrivial dependence on the worldsheet
$\s$-coordinate (see \eq{X0})
and is  
more complicated than  the case without $B$-field.
Also we find that the zero mode frequency is affected by a
constant background $B$-field,
but not for the higher oscillation modes. 
In section 3.2, we review the procedure for determining the commutation
relations of the theory. The starting point is the symplectic form
\eq{omega}. We explain why it is consistent and use it to derive the
basic commutation relations \eq{zCR1}--\eq{zCR4} 
for the string modes.
As a result of these commutation relations, the boundary commutators
of the string coordinates and string canonical momentum are modified,
and only at the boundary, as in \eq{XX1}--\eq{XP1}. 
The spectrum of the lightcone Hamiltonian is obtained. As a consequence of these 
commutation relations,  
we find that the  spectrum of the nonzero modes depend on 
the mode number $n$ in a more complicated manner; we also noted 
an interesting  splitting in the zero mode spectrum, see \eq{lcH2}.
In section 3.3, we carry out the constrained quantization of Dirac and
obtain the same result.
In section 4, we discuss various scaling limits and discuss their
physical meanings. 
We conclude the paper with  further discussions.

\section{BPS D-brane with $B$-field in pp-wave}

Consider a flat D-brane in the pp-wave background, 
with a constant $B$-field turned on in its worldvolume. 
This is a BPS configuration preserving half of the 32 pp-wave supersymmetries.
This can be easily seen 
from the limiting procedure
considered in \cite{bfhp2,bfhp3} that relate brane probe solution in $AdS
\times S$ to brane probe solution in the Penrose-limited pp-wave or
Minkowskian background. 
In this section, we will use  
the $\k$-symmetric formulation of D$p$-brane (see for example 
\cite{schwarz,cederwall,d1}) 
to show this explicitly.
It will also allow us to 
see how the D-brane is stuck at the origin from the point of view of
the Born-Infeld theory, see \eq{lift}.

Given a D-brane probe, the surviving supersymmetries satisfies
the condition 
\be \label{cond1}
(1-\C) \xi =0
\ee
where $\C$ is a projector that depends on the details  of the
brane configuration. 
The explicit form of $\C$ was obtained, 
for example,  in \cite{bkog}, where it
was shown that the Born-Infeld field strength amounts to a relative
rotation of the left and right moving fields. In order to be self
contained, we recall that $\C$ is given by
\be \label{aga}
\C = e^{-a/2} \C_{(0)}' e^{a/2},
\ee
where $a$ is a matrix given below and
\be \label{gamma0'}
\C'_{(0)}=\cases{ (\C_{11})^{p-2\over 2} \C_{(0)}& IIA,\cr
(\s_3)^{p-3\over 2} i \s_2\otimes \C_{(0)}\ &IIB,}
\ee
and
\be \label{gamma0}
\C_{(0)}= \frac{1}{(p+1)! \sqrt{-\det G}} \epsilon^{i_1\cdots i_{p+1}}
\del_{i_1} X^{M_1} \cdots \del_{i_{p+1}}X^{M_{p+1}}
\C'_{M_1 \cdots M_{p+1}}.
\ee
As usual, the matrix  $\C'_{M_1 \cdots M_{p+1}}$ is the
antisymmetrized product of the $\C_{M_k}'$ with the $\C_{M}'$
being the 10 dimensional $\C$-matrices in the
coordinate basis defined by
\be
\C'_M := E_M{}^A \C_A,
\ee
where the $\C_A$ are flat space $\C$-matrices.
The metric $G_{ij}$ is the induced worldvolume metric
\be
G_{ij} = \del_i X^M \del_j X^N g_{MN}.
\ee
To define the matrix $a$ appearing in \eq{aga} we need to introduce
the modified 2-form field strength $\cF$ which is related to the Born-Infeld
field strength $F= dA$ by
\be
\cF =F - \underline{B},
\ee
where $\underline{B}$ is the pullback of the target space NS-NS
2-form potential to the worldvolume.
The matrix $a$ depends only on the worldvolume Born-Infeld field
strength and is given by
\be \label{a}
a = \cases{-{1\over2} Y_{jk} \c^{jk} \C_{11} & IIA, \cr
            {1\over2} Y_{jk} \s_3\otimes\c^{jk}&IIB,}
\ee
the $\c^{jk}$ being worldvolume $\c$ matrices,
\be
\c_i = \del_i X^M \C'_M
\ee
and  $Y$ is a function of $\cF$. The relation in the frame basis of
the worldvolume (hatted indices) is
\be
Y_{\hat{j}\hat{k}}:=
\mbox{tan}^{-1} \cF_{\hat{j}\hat{k}}.
\ee

Specifically, let us consider a D$p$-brane spanning the directions 
$(+,-, i_2, \cdots, i_p)$,
where $X^{\pm}= (X^0\pm X^9)/\sqrt{2}$ and $(i_2,\cdots,i_p) \in (1,\cdots,8)$. 
Denote the worldvolume coordinates of the D-brane by
$(\t, \xi^k), k=1,\cdots,p$. In the lightcone gauge,
\be
X^+(\t,\xi) = p^+ \t.
\ee
We also take the physical gauge
\be
X^-(\t,\xi)= \xi^1, \quad\quad
X^{i_k}(\t,\xi) = \xi^k, \quad k= 2,\cdots, p.
\ee
The transverse directions $X^a(\t,\xi)$, 
then describe the embedding of the D-brane.

We will consider 
a constant $B$-field with nonvanishing components $B_{ij}$, 
$i,j \in (1,\cdots,8)$.
The pull-back $B$-field is the same as the target space
components, and therefore we will not distinguish them anymore.
As for the pullback metric, we have  
\be \label{phy-metric}
G_{\m\n} = G^{(0)}_{\m\n} +  \del_\m X^a \del_\n X^a,
\ee
where 
$G^{(0)}_{\m\n}$ is the metric 
\be
G^{(0)} = \pmatrix{
-u^2 & p^+ & \cr
p^+ & 0 & \cr
 & & \uno_{(p-1)\times (p-1)}
}
, \quad 
u^2 := p^{+ 2}\,{\rm f}^2 (\sum_{i=2}^p \xi^{i\,2} + X^{a\,2}).
\ee
Now, it is easy to see that for a flat D-brane described by 
\be \label{Xconst}
X^a(\t,\xi) =  \mbox{const},     
\ee
we have
\be \label{c0}
\C_{(0)} = \C_{+- i_2 \cdots i_p}.
\ee
This is  independent of $B$ and $m$ and the  D-brane preserves half  
supersymmetries.
Using the form of the killing spinor given in \cite{bfhp1} and the fact that
the brane is stuck at the origin (see \eq{lift} below), it is easy to check
that $\C_{(0)}'$ preserve half of the supersymmetry for
$p=3,5,7$. This agrees with the open string analysis of \cite{dab} \footnote{We 
thank Atish Dabholkar, Jaume Gomis 
and Shahrokh Parvizi for useful email exchanges.}. 
Branes at angle \cite{angles} can be discussed similarly. 

Finally we check that the brane configuration satisfies the equations of
motion.
In our case, since we are using the physical gauge,
the  equations of motion are
derived from the gauge fixed form of the Dirac-Born-Infeld
action, including the WZ-term,
\be \label{dbi}
I = \int d^{p+1} \xi \sqrt{-{\rm det} (G + \cF) } + \int \underline{C}
\; := I_{DBI} + I_{WZ},
\ee
where $G_{\m\n}$ is the induced worldvolume metric  \eq{phy-metric} 
in the physical gauge and
$\underline{C}$ is the pullback to the worldvolume of the RR
4-form potential. The inclusion of the WZ term is crucial in checking
that the equation of motion is satisfied for brane
configurations with nontrivial embeddings \cite{BCk}. 
However for the trivial  embedding \eq{Xconst}, the WZ term does
not play any role.
The equation of motion reads
\bea 
& \Delta^{-1}\del_\m \left(\Delta
\{(G+\cF)^{-1} +(G-\cF)^{-1}\}^{\m\n}\del_\n X^a \right) -
m^2 X^a =0, \label{e1}\\
& \del_\m \left(\Delta
\{(G+ \cF)^{-1} -(G - \cF)^{-1}\}^{\m\n}\right) =0, \label{e2}
\eea
where
\be
\Delta := \sqrt{-\det (G + \cF )} .
\ee
The only constant solution of equation \eq{e1} is
\be \label{lift}
X^a = 0.
\ee
The absence of flat directions is because they are lifted by the 
potential well created by the mass term.
Equation \eq{lift} implies that
\be
u^2 = p^{+ 2}\,{\rm f}^2 \sum_{i=2}^p \xi^{i\,2}.
\ee
Note that it is independent of neither $\t$ nor $\xi^1$.
It is then easy to check that 
\be
\cF_{ij}= \mbox{const}
\ee
is a solution to \eq{e2}. Hence our claim is justified.
In the following, we will turn to the open
string description for these D-branes. For simplicity, we will take 
the gauge $\cF =B$ where $A=0$.

\section{Noncommutative D-brane  with $B$-field in pp-wave}
 
The main objective of this paper is to work out the
effects of $B$-field on the quantization of the 
open string theory ending on a D-brane  in the 
pp-wave background. We will carry out this analysis in the GS
formulation. 

\subsection{Open String Mode Expansion}

Recall that for  a flat D-brane in Minkowskian spacetime, the open string  
GS action  was
obtained from lightcone gauge fixing the covariant action \cite{GG},
in which the latter can be obtained by substituting in the
superfield background that solves the IIB supergravity constraints. 
With  a constant $B_{\m\n}$ field turned on, it 
induces new nonzero components in the superfields $B_{\m\a}$ and $B_{\a\b}$.
However these terms all cancel themselves out in the action and  
the only effect  is the addition of 
the usual bosonic $B$-field coupling to the target
space \cite{CZ}. A coupling to the fermionic spacetime
variables would arise only if $B$ were not closed.
It was also shown that one can use the usual (ie. the one for $B=0$) 
lightcone gauge fixing condition so long as $B_{0 \m}= B_{9 \m} =0$
\cite{CZ}. This is exactly the setting we considered in 
section 2.

The $\k$-symmetric formulation of closed string in pp-wave
background was analysed in \cite{metsaev}. In the lightcone gauge,
the theory is exactly solvable and consists of eight massive bosons
and fermions. Turing on the $B$-field, and carrying out a similar analysis
as in \cite{metsaev,CZ}, one finds that
the open string has the bosonic action in the lightcone gauge,
\be \label{S} 
S = \frac{1}{4\pi \a'} \int d^2 \s [ 
g_{ij} (\eta^{\a\b}\del_\a X^i \del_\b X^j +m^2 X^i X^j)
+ \e^{\a\b} \del_\a X^i \del_\b X^j B_{ij} ],
\ee
where $i,j = 1, \cdots p$,
and $B$ is turned on only in the directions $2, \cdots, p$.
Following \cite{t2}, we have introduced the parameter
\be \label{mass}
m : = \a' p^+ {\rm f},
\ee
and we have taken  the length of the $\s$-interval  to be 
$\pi$; therefore strings with different lightcone momentum have different $m$. 
We ignored the transverse coordinates in \eq{S} since the
quantization of them is standard. The fermionic sector will not affect
the result and can be considered separately. We have introduced the
Euclidean metric $g_{ij}$ 
for the sake of the scaling limits to be considered in the next section.
Note that in our normalization, $B, m$ are dimensionless.
The equation of motion is 
\be \label{eom}
(-\del_\t^2 + \del_\s^2 -m^2) X^i =0, 
\ee
and the boundary condition is
\be \label{bc}
\del_\s X^i + \del_\t X^j B_j{}^i =0
\ee
at $\s=0,\pi$.  Indices $i,j$ are raised and lowered by $g_{ij}$.  
The lightcone Hamiltonian is \cite{metsaev,t1}
\be \label{lcH}
H =\frac{1}{4\pi \a' p^+}\int_0^{\pi} d \s
\left( (\del_{\t}X)^2+(\del_{\s}X)^2 +m^2 X^2
\right),
\ee

We are interested in the effects of the constraint \eq{bc}
to the quantization of the theory.  
The situation is exactly the same as in \cite{CH1}. The boundary
condition  \eq{bc} implies that
\bea\label{PPcr}
&2\pi\a'[P^k(\t,0),P^j(\t,\s')] B_k{}^i=
-\del_{\s} [X^k(\t,\s),P^j(\t,\s')]_{\s=0} \, M_k{}^i, \\
&2\pi\a'[P^k(\t,0),X^j(\t,\s')] B_k{}^i =
- \del_\s [X^k(\t,\s), X^j(\t,\s')]_{\s=0} \, M_k{}^i,
\eea
where $P^k$ is the conjugate momentum 
\be \label{mom}
2\pi\a' P^k(\t,\s)=\del_\tau X^k + \del_\sigma X^j B_j{}^k,
\ee
and
\be \label{M}
M_k{}^i = \delta_k{}^i - B_k{}^j B_j{}^i.
\ee
These simple relations show that the standard canonical commutation
relations for $B=0$ \cite{metsaev} are no longer valid when $B \neq 0$. 

Without loss of generality, we will assume the metric is already in the
diagonal form
\be \label{diag-g}
g_{ij} = \l \delta_{ij}. 
\ee
In addition we will consider the case in which the $B$-field takes the form
\be \label{B12}
B_{ij} = \pmatrix{0 & \B \cr -\B & 0},
\ee
and focus on $X^2, X^3$.
We remark that the action \eq{S} is $SO(2)$ invariant,
and so the angular momentum is conserved. 
The mode expansion takes the form
\be \label{X}
X^i = X_{(0)}^i + X_{(1)}^i,
\ee
where 
\footnote{
It is straightforward to extend the two dimensional case
to the generic case. Assuming that $B_{\m\n}$ is invertible,
we can always decompose it
as $U^T b U$ by an orthogonal transformation $U$, where
$b$ is a matrix of diagonal blocks of the form (\ref{B12}).
Then the zero mode solution is
\be
X_{(0)}^i = x_+^j {\left( U^T \exp(+(ab^{-1}\tau-a\s)) U \right)_j}^i
          + x_-^j {\left( U^T \exp(-(ab^{-1}\tau-a\s)) U \right)_j}^i,
\ee
where the matrix $a$ is a diagonal matrix defined by
\be
a^2 = \frac{m^2}{1-b^{-2}}.
\ee
Note that $a$ is well defined by this relation
because $b^{-2}$ is diagonal with negative entries,
}
\bea  \label{X0}
X_{(0)}^i=  
&&(x_0^i \cos \o_0 \t + 2 \a' p_0^i\frac{\sin \o_0 \t}{\o_0} ) \cosh \o_0 \bB \s
\nn\\
&&+ ( -  2 \a' p_0^j  \cos \o_0 \t  +  x_0^j \o_0 \sin \o_0 \t ) B_j{}^i
\frac{\sinh \o_0 \bB\s}{\o_0 \bB},
\eea
is the ``zero mode'' part,
and
\be \label{X1}
X_{(1)}^k = \sqrt{2 \a'}\sum_{n \neq 0} e^{- i \o_n \t} (i \,\frac{ \a_n^k}{\o_n}
\cos n \s - \frac{\a_n^j}{n} B_j{}^k \sin n \s),
\ee
is the nonzero mode part.
The constant $\bB$ 
\be
\bB := B/\l,
\ee
is the eigenvalue of the matrix $B_i{}^j$; and the frequencies are defined by
\be
\o_0 := \frac{m}{\sqrt{1+\bB^2}} >0, \quad \quad \mbox{and} \quad
\o_n := {\rm sign}(n) \sqrt{n^2+m^2}, \quad n\neq 0.
\ee
A couple of remarks about the ``zero modes'' are in order. 
{\bf 1.}  By ``zero modes'' here, we simply refer to the modes with the lowest
frequency.
We stress that the zero modes are very different in structure compared
to the oscillation modes. One important difference between the zero mode part 
and the oscillation part is that the zero mode part $X_{(0)}^i$
actually satisfies
\eq{eom} and \eq{bc} for all $0 \leq \s
\leq \pi$. This is the same as in the flat case ($m=0$) \cite{CH1}.
{\bf 2.} Note that the zero mode
structure is  more complicated than the $B=0$ case,
where $X^i_{(0)}$ was simply independent of $\s$ \cite{metsaev}. 
{\bf 3.} In our expansion \eq{X0},
the coefficients $x_0$ and $p_0$  are identified   
conveniently such that they have the correct dimensions.
It is a simple choice and it is possible to rescale them by
dimensionless factors.   
{\bf 4.}
Finally, we note that
naively one may have expected that, since turning on a constant
$B$-field has no effect on the equation of motion (propagation), 
the zero mode will have the frequency $m$ as in the $B=0$ case
\cite{metsaev}. 
However, it is easy to check that when $B \neq 0$, 
this frequency does not give rise to any
solution that can satisfy both the equation of motion
and the boundary condition.
The modification of the zero mode frequency by the external $B$-field
through the boundary condition is an interesting effect.
We recall that in the flat case for a neutral string, 
turning on a B-field does not change the mode frequency. In the case
of a charged string \cite{callan}, all the frequencies are modified by
the background field. Our case here is intermediate, turning on a
$B$-field only modify the zero mode frequency. 
This is another
difference between the zero modes and the oscillation modes.

The canonical momentum also splits into a sum 
\be \label{P}
P^k = P_{(0)}^k + P_{(1)}^k
\ee 
of the the zero mode part
\bea \label{P0}
2 \pi \a' P_{(0)}^k =
&&  
( -  x_0^j\o_0 \sin \o_0 \t + 2 \a' p_0^j\cos \o_0 \t )
M_j{}^k \cosh \o_0 \bB \s
 \nn\\ 
&&+ ( 2 \a' p_0^j \sin \o_0 \t  +  x_0^j \o_0  \cos \o_0 \t )
(Bg^{-1}M)_j{}^k \;
\frac{\sinh \o_0 \bB\s}{\bB},
\eea
and the nonzero mode part 
\be \label{P1}
2 \pi \a' P_{(1)}^k =\sqrt{2 \a'} \sum_{n \neq 0} e^{-i \o_n \t} 
(\a_n^j M_j{}^k \cos n \s + i \frac{m^2}{n \o_n} \a_n^j B_j{}^k\sin n\s).
\ee
Note that \eq{X0}, \eq{X1}, \eq{P1} and \eq{P0}
reduce smoothly to the usual flat
space expressions in the limit $m \to 0$.

\subsection{Open String Quantization in pp-wave with B-field}

To quantize the theory we will follow the procedure 
adopted in \cite{CH1,CH2}, 
where interpretations in terms of 
noncommutative geometry were emphasised.
This method is equivalent to the 
canonical quantization performed in \cite{callan}. 
That
paper also gave the first instance demonstrating how 
noncommutativity of the zero modes can arise in the presence of 
a background field. Its physical implications 
had been studied in \cite{callan,bachas}. 

First we need the symplectic form 
\be\label{omega}
\Omega=  \int_0^\pi d\s g_{ij} \bd P^{i} \bd{X}^{j}.
\ee 
That this is the correct symplectic form of the theory 
can be justified from an analysis of the constraint structure of
the theory. Without going into the details, one can already see this
since the constraint \eq{bc} is imposed only at the
boundary. It is clear that the  constrained quantization method
will give a Dirac bracket which is modified and is possibly
different from the original Poisson bracket only at the boundary. 
(We will present the constrained quantization
in the next subsection.)
This is a measure zero set and so the symplectic form takes the
standard form \eq{omega}. 
 
Of course this doesn't mean that the 
commutation relations of the theory are unmodified. 
This is felt through the modified form of the mode expansion \eq{X0},
\eq{X1}, \eq{P0}and \eq{P1}.  
By substituting in the mode expansions \eq{X}
and \eq{P} and evaluate the $\s$ integral, the
symplectic form \eq{omega} of the theory can  be thought of as the
symplectic form defined for the string modes.  This procedure is
consistent since, using \eq{eom} and \eq{bc}, it is easy to check that
\eq{omega} is independent of $\t$ \cite{neutral}.
As a result, one can 
take the resulting expression as the symplectic form for the string modes. 
The commutation relation of the string modes can be
then obtained from \eq{omega} by inverting the symplectic matrix.

Explicitly we obtain the following symplectic form for the string modes 
\bea
2 \pi \a' \Omega = && 
2 \a' \ct \; M_{ij}\bd p_0^i \bd x_0^j   - 
\frac{\o_0 c}{2 \bB} (Mg^{-1}B)_{ij} \bd x_0^i \bd x_0^j
- (2\a')^2 \frac{ c}{2 \o_0 \bB} (Mg^{-1}B)_{ij} \bd p_0^i \bd p_0^j
\nn\\
&&+i 2\pi \a'  \sum_{n=1}^{\infty} \frac{M^n_{ij}}{\o_n} \bd \a_n^i \bd \a_{-n}^j,
\eea
where the ``metric'' $M_{ij}^n$ is defined by
\be
M_{ij}^n := g_{ij} - \frac{\o_n^2}{n^2} (B g^{-1} B)_{ij}.
\ee
Note that $M^n_{ij}= M_{ij}$ only when $m=0$. 
The constants $c, \ct$ are defined by
\bea
c &=& (\cosh (2 \pi \o_0 \bB) -1)/(2 \o_0 \bB) \approx \pi^2 \o_0 \bB 
+ \cdots, \\
\ct &=& \sinh (2 \pi \o_0 \bB) /(2 \o_0 \bB)\approx \pi + \cdots.
\eea
The $ \approx$ above gives the leading order expansion in the small $m$ limit. 
Following \cite{SW}. we have been careful in keeping the metric
dependence to facilitate the discussion of scaling limits. This will be
done in the next section.

Inverting the symplectic matrix, we obtain
the following commutation relations for the zero modes and for the
oscillation modes
\bea 
&&[x_0^i, p_0^j] = i  (M^{-1})^{ij} \;
\frac{\pi\o_0 \bB}{\tanh\pi\o_0 \bB }  , \label{zCR1}\\
&&[x_0^i,x_0^j] = i 2 \pi \a' (g^{-1}B M^{-1})^{ij}   , \label{zCR2}\\
&&[p_0^i, p_0^j] = i \frac{\pi \o_0^2}{2 \a'} (g^{-1}B M^{-1})^{ij}, \label{zCR3}
\eea
\be \label{zCR4}
[\a_n^i, \a_s^j] = \o_n  M_n^{ij}\d_{n+s},
\ee
where
\be
M_n^{ij} = ( \frac{1}{g+ \frac{\o_n}{n} B} \;g\; \frac{1}{g- \frac{\o_n}{n} B})^{ij}
\ee
is the inverse of $M^n_{ij}$  for each fixed $n$.
It represents a metric that is mode-dependent.
This is in contrast with the flat case, where it was 
found that  \cite{CH1} the oscillator's commutation relations are
determined by the same open string metric $M^{ij}$ as the zero
modes. Here we find that each level of the oscillators see a different
metric. 
A consequency of this is  the more complicated $n$-dependence in the 
spectrum of the Hamiltonian \eq{lcH2}.
It may be  interesting to investigate to what extent one can think
of $M_n^{ij}$ as a mode-dependent open-string metric.
Note that in the form  \eq{diag-g} of the metric, 
\be
M_n^{ij} = g^{ij}(1+\frac{\o_n^2}{n^2} \bB^2)^{-1} 
\ee
is diagonal and is a monotonic increasing function of $n$
which is  bounded above by $M^{ij}$. 
Note also that  by turning on the mass parameter $m$, $p_0^i$ also become
noncommuting themselves; and the phase space of  $x_0, p_0$
becomes fully noncommuting. The noncommutative parameters depends on
the background of the pp-wave, and on the lightcone momentum $p^+$.

In the above, we have considered a single string with a single
$p^+$.  If one wants to perform calculations with many string states, then
it will be useful
to introduce a basis of oscillators whose commutation
relations are independent of $p^+$. For the nonzero modes, we can
introduce the following basis of oscillators 
\be
a_n^i := \sqrt{ \frac{1+\frac{\o_n^2}{n^2} \bB^2}{\o_n/\l}} \a_{-n}^i, \quad 
\bar{a}_n^i := \sqrt{\frac{1+\frac{\o_n^2}{n^2} \bB^2}{\o_n/\l}} \a_{n}^i,
\quad n > 0.
\ee
They obey the relations
\be
[\bar{a}_n^i, a_s^j] = \d^{ij} \d_{ns}\, , \quad[a_n^i, a_s^j] = 0. \label{a1}
\ee
Note the $B$-dependence in the rescaling. 
For the zero mode, one may define
\be 
a_0^i :=  \frac{w}{2}
(x_0^i - i \frac{ 2 \a' p_0^i}{\o_0}), \quad
\bar{a}_0^i := (a_0^i)^\dag, 
\quad \mbox{with}\quad
w:= 
\sqrt{\frac{\l m^2}{ \a' \o_0} \frac{\tanh \pi \bB \o_0}{\pi \bB\o_0}},
\ee
in terms of which the commutation relations \eq{zCR1}--\eq{zCR3} take
the simple form of  a deformed oscillator algebra
\be \label{a20}
[\bar{a}_0^i, a_0^j] = \d^{ij} + i \e^{ij}\cdot \tanh \pi \bB \o_0\, ,
\quad [{a}_0^i, a_0^j]=0,
\ee 
which is $SO(2)$ invariant. Or one may introduce
\be
a_{\pm} := \frac{a^1_0 \pm i a^2_0}{\sqrt{2(1\mp\tanh\pi\bar{B}\o_0})},
\quad \bar{a}_{\pm} := (a_{\pm})^{\dagger},
\ee
which obey the $U(1)$ invariant commutation relations
\be \label{a2}
[\bar{a}_A, a_B] = \delta_{AB}, \quad [a_A, a_B] = 0, \quad \quad\quad A, B = +, -.
\ee

In terms of them, the zero mode part of the
string coordinates can be written as, for example, 
\be
X_{(0)}^i = \frac{e^{i \o_0 \t} }{w} 
(a_0^i \cosh \o_0 \bB \s - i a_0^j B_j{}^i \frac{\sinh \o \bB
\s}{\bB}) + {\rm h.c.} \, .
\ee
The lightcone Hamiltonian \eq{lcH} takes the diagonal form
\bea \label{lcH2}
H &=& \frac{\o_0\cosh^2 \pi \bB \o_0 }{ p^+}  
\left( (1-\tanh\pi\bB\o_0)a_+\bar{a}_+ +
(1+\tanh\pi\bB\o_0)a_-\bar{a}_- + 1 \right) \nonumber \\
&&+ \frac{1}{p^+} \sum_{n=1}^\infty 
\; \frac{1+ \bB^2}{1+  \frac{\o_n^2}{n^2}\bB^2 } \; \o_n   a^i_n \bar{a}^i_n \, ,
\eea
where we have written the Hamiltonian in a normal ordered form, and
have dropped the infinite additive constant arising from the nonzero modes.
This will be cancelled with the fermionic oscillators \cite{t2}. The factor 1
in \eq{lcH2} came from the normal ordering of the zero mode oscillator
$a_{\pm}$ and remember that we are considering two string coordinates $X^{2,3}$.

The  Fock vacuum of the theory is defined by
\be
\bar{a}_\pm \ket{0} =0, \quad \bar{a}_n^i \ket{0} =0,\; n>0. 
\ee 
It is $SO(2)$ invariant. In choosing the definition of the vacuum, we
have been guided by the requirement of $SO(2)$ invariance of the
theory. Turning on a $B$-field in the 2--3 directions preserve this symmetry.
Note that the spectrum for the nonzero modes depend 
on $n$ in a more complicated way due to $B \neq 0$.
Note also that the degeneracy of the states  $a_\pm \ket{0}$ 
is lifted when $B \neq 0$. 
We expect that this  splitting of the string spectrum to have interesting 
physical consequences. 
 
Finally we  derive the intrinsic commutation relations of the
theory in terms of $X^i$ and $P^i$. 
Using the relations above, one can easily obtain
\be \label{XX}
[X^i(\t,\s), X^j(\t,\s')] = i 2\pi \a' \B^{ij} f(\s+ \s'),
\ee
where the function $f(\s)$ is defined on $\sigma  \in  [0,2\pi]$ as
\be
f(\s) := \frac{1}{c}[c \cosh (\o_0 \bB \s) - \ct \sinh (\o_0 \bB \s) ]
- \frac{2}{\pi} \sum_{n=1}^{\infty} 
\frac{1}{n}(1+\frac{\o_0^2 \bB^2}{n^2})^{-1} \sin n \s.
\ee
It is easy to show that
\be
f(\s) =(1+\bB^2)^{-1} \times
\left\{ 
\begin{array}{cl} 
+1, & \s =0 ,\cr
-1 , & \s = 2 \pi,\cr
0, & 0 < \s  < 2\pi.
\end{array}
\right.
\ee
Thus the commutation relation of $X$ is
\be \label{XX1}
[X^i(\t,\s), X^j(\t,\s')] = i 2\pi \a' (g^{-1}BM^{-1})^{ij} \times
\left\{ 
\begin{array}{cl} 
1 , & \s = \s' =0 ,\cr
-1 , & \s =\s' =  \pi, \cr
0, & \mbox{otherwise}. 
\end{array}
\right.
\ee 
and is  modified only at
the endpoints of the open string. 
The commutator for $P$ themselves can be
similarly computed and we obtain
$$
[P^i(\t,\s), P^j(\t,\s')] = i \frac{m^2}{2\pi \a'} (1+\bB^2) \B^{ij}
f(\s+ \s'). 
$$
Thus both  commutators 
are given in
terms of the same function $f$. As a result, we find
\be \label{PP1}
[P^i(\t,\s), P^j(\t,\s')] = i \frac{m^2}{2\pi \a'} B^{ij} \times\left\{ 
\begin{array}{cl} 
+1 \;, & \s =\s'=0 ,\cr
-1 \;, & \s = \s'= \pi, \cr
0, \;& \mbox{otherwise}. 
\end{array}
\right.
\ee
It is intriguing that the commutation relations
for the endpoint momentum are nontrivial.
Note also it dependent on the parameter
$m = \a' p^+ {\rm f}$
in a nontrivial way. 
This is a new feature of the pp-wave background:
the noncommutative space felt by a string
depends on its light-cone momentum.
Finally we obtain
\be \label{XP1}
[X^i(\t,\s), P^j(\t,\s')] = i \,\frac{g^{ij}}{\pi}  (1+ \sum_{n \neq 0} \cos
n \s \cos n \s').
\ee
Note that \eq{XX1} and \eq{XP1}  take exactly the same form
as in the flat case \cite{CH1} and is unmodified by $m$. 
Also  note that, unlike the relations for the zero modes, 
the relations  \eq{XX1}--\eq{XP1} are intrinsic and is
independent of how the zero modes are identified, nor how the vacuum
is chosen.

\subsection{Constrained Quantization}

In this subsection, we present the construction of the
Dirac bracket following the procedure in \cite{CH2}.
To start with, the standard Poisson bracket is \cite{metsaev}
\bea 
(X^i(\s),P_j(\s'))=\d^i_j\d(\s,\s'), \quad
(P_i(\s),P_j(\s'))=0, \quad
(X^i(\s),X^j(\s'))=0.
\eea
As first noted in \cite{CH1}, the boundary condition can be treated as
a constraint on the phase space
\be
\Phi^i( 0) = \Phi^i(\pi) =0, 
\ee
where we have introduced 
\be
\Phi^i (\s) :=
2 \pi \a' P^j B_j{}^i +  \del_\s X^j M_j{}^i.
\ee
Using the lightcone Hamiltonian \eq{lcH}
one determines the complete set of second class constraints
\be
\del^{2\iin}_\s \Phi^i(\s)=0, \quad\quad \del^{2\iin}_\s \Psi^i (\s) =0,
\quad\quad  \iin=0,1, \cdots ,
\ee
where 
\be
\Psi^i (\s):= \del_\s P^i - \frac{m^2}{2 \pi \a'}  X^j B_j{}^i.
\ee
We will denote
them by $\phi^{(\a k \iin)}$, $\a =1,2$; $k=2,\cdots,p $;
$\iin =0,1, \cdots $,
\be \label{allcons}
\phi^{(1k \iin)} = \del^{2\iin}_\s \Phi^k, \quad
\phi^{(2k \iin)} = \del^{2\iin}_\s \Psi^k.
\ee
These constraints are consistent with the 
explicit mode expansion of the fields $X^i$ and $P^i$ given above.
 
The Poisson matrix $C^{(\a k \iin) (\b l \iim)}$
of the constraints can be computed easily. The basic ones are the
$C^{(\a k  0) (\b l 0)} \; $ 's :
\bea
(\Phi^k(\s), \Phi^l(\s')) &=& 2\pi\a' (B M)^{kl}
[\del_\s \d(\s,\s') + \del_{\s'} \d(\s',\s) ], \label{pb1} \\
(\Phi^k(\s), \Psi^l(\s')) &=&
M^{kl} \del_\s \del_{\s'} \d(\s,\s') + m^2 (B^2)^{kl}
\d(\s,\s') , \label{pb2}\\
( \Psi^k(\s), \Psi^l(\s')) &=&  \frac{m^2}{2 \pi \a'} B^{kl} 
[\del_\s \d(\s,\s') + \del_{\s'} \d(\s',\s)] \label{pb3}
\eea
and in general
\be
C^{(\a k \iin) (\b l \iim)}(\s,\s')
=(\phi^{(\a k\iin)}(\s), \phi^{(\b l\iim)}(\s'))
=\del_\s^{2\iin} \del_{\s'}^{2\iim} C^{(\a k 0) (\b l 0)}(\s,\s').
\ee
The Dirac bracket is given by
\bea \label{DB}
(A(\s), B(\s'))^* &=&(A(\s), B(\s')) \nn \\
&-& \sum_{\s^{''} \s^{'''}} (A(\s), \phi^{(\a k \iin)} (\s^{''}))
C_{(\a k \iin) (\b l \iim)}(\s^{''}, \s^{'''} )
(\phi^{(\b l \iim)} (\s^{'''}), B(\s')), \nn \\
&&
\eea
where the sum  $\s'', \s'''$ is over the endpoints $0, \pi$.
As in \cite{CH2}, the Dirac bracket can be computed similarly without knowing
the explicit form of the inverse matrix $C_{(\a k \iin) (\b l \iim)}$. Only
its defining properties are needed. We find the results 
\eq{XX1}--\eq{XP1}.
In particular we note that the $\s,\s'$ dependent part of
the right hand side of \eq{pb1} and
\eq{pb3} are identical and this accounts for the fact that the
commutators \eq{XX1} and \eq{PP1} are both given by the same
function $f(\s+\s')$.  

\section{Scaling Limits}

In this section, we will consider scaling limits of $\a', B, m$ 
such that the commutation relations (\ref{zCR1})--(\ref{zCR3})
remain nontrivial. 
In these limits the noncommutative algebra
should play a role in the D-brane physics.
We remark that one can equivalently perform the limits on \eq{a2}
for the variables $a_0^i$'s. Here we choose to consider the variables
$x_0^i, p_0^i$ since we would like to explore and develop 
possible geometrical interpretations.

\subsection{Small $B$-Field Limit}

The DBI action is exact to all orders in $\a'$ \cite{tseytlin}
for the $U(1)$ gauge field ${\cal F}$. 
Here we prefer to think of it as a consequence of
supersymmetry (see, for example \cite{schwarz,cederwall,d1}).  
For small $B$-field, it should be possible to
check that the DBI action on pp-wave background
is equivalent to the DBI action
on the noncommutative space without $B$-field
for slowly varying fields,
in a similar way as it was checked for the flat
background \cite{SW}.
This match may provide some hint for the
appropriate description of the noncommutative space.
A particularly interesting limit is 
to go to the infinite momentum
frame and take a small $B$-field simultaneously:
\be
B \sim \e^2, \quad m \sim 1/\e, \quad \mbox{$g_{ij}, \; \a'$ fixed}.
\ee
The commutation relations (\ref{zCR1})--(\ref{zCR3})
become
\bea
{[x_0^i, p_0^j]} &=& i g^{ij}, \label{bCR1} \\
{[x_0^i, x_0^j]} &=& i2\pi\a' B^{ij}, \label{bCR2} \\
{[p_0^i, p_0^j]} &=& i \frac{\pi  m^2}{ 2\a'} B^{ij}. \label{bCR3}
\eea
In this limit the coordinates noncommutativity
can be treated perturbatively, but
the description of the momentum noncommutativity has to be
exact. 
This limit is interesting since the $p^+$ dependence has
essentially disappeared because one can treat all the strings to have the
same lightcone momentum in the leading order approximation. It would
be interesting to study the corresponding large $N$ matrix model.

\subsection{Large $B$-Field Limit: Deformed Phase Space}

When $m = 0$, all the excited modes can be ignored
in the limit $\a' \rightarrow 0$. And we only need to
consider the zero modes for the low energy effective theory. Moreover,
if we also take the large $B$-field limit \cite{SW} 
of  Seiberg and Witten :
\be \label{limit2}
\a' \sim \e^{1/2}, \quad g_{ij} \sim \e, \quad \bB \sim \e^{-1/2},
\ee
then a noncommutative theory is obtained with
$M$ (\ref{M}) and the noncommutativity parameter
\be \label{limit-theta}
\theta^{ij}:=2 \pi \a' (g^{-1}B M^{-1})^{ij}
= \frac{2 \pi b^{ij}}{b^2}
\ee
fixed and finite in the limit. 
To facilitate the comparison, our $B_{ij}$ is related to the
dimensionful $b_{ij}$ of Seiberg-Witten \cite{SW}
by $B_{ij} = \a' b_{ij}$. Hence
$\bB = \a' b /\l \sim \e^{-1/2}$ as $b$ is fixed.

However, for $m \neq 0$, the zero mode frequency $\o_0$ is of the same
order of magnitude as the oscillator modes,
and the ratio $\o_0 / \o_1$ is finite in the limit
$\a' \rightarrow 0$.
Therefore generally we cannot ignore the higher frequency modes
unless we take another suitable limit at the same time.
It is easy to see that for fixed $m$, the Seiberg-Witten limit
\eq{limit2} does the job. It sends $\o_0 \sim \e^{1/2}$ while keeping
$\o_{n \neq 0}$ fixed. The zero
modes commutation relations take the form
\bea
&&[x_0^i, p_0^j] = i\, h^{ij} , \label{fCR1}\\
&&[x_0^i,x_0^j] = i\, \th^{ij}, \label{fCR2}\\
&&[p_0^i, p_0^j] = i\,  \,\k^{ij}. \label{fCR3}
\eea
where $\th^{ij}$ is given by \eq{limit-theta} and
\be 
h^{ij}:= \frac{\lambda}{\a'^2 b^2}
\;\frac{\pi m}{\tanh\pi m } \delta^{ij},
\quad \k^{ij}:=  \frac{\lambda^2 m^2}{4 \a'^4 b^2} \th^{ij}
\ee
are fixed and finite in the limit.  Note that $\th$ has dimension
(length)${}^2$ and $\k$ has dimension (length)${}^{-2}$. Note also
that the reason that $\k \propto \th$ is because here we are dealing with
2-dimensions only. This will not be the case if we consider a higher
rank $B$-field.

The commutation relations
\eq{fCR1}--\eq{fCR3} are intriguing. They represent a phase space
with a fully noncommutative structure. 
If $m$ is zero, we 
go back to the usual case of a
theory defined on a noncommutative manifold.
For $m \neq 0$, a
deformed phase space
emerges with the
noncommutativity parameters $\th^{ij}$ and $\k^{ij}$. Moreover these
parameters depends on the pp-wave background and the string momentum. 
It will be very interesting to understand the implications of this. 

As a simpler question, one can momentarily forget
about the string embedding and 
consider simply the physics of the noncommutative phase space by
itself.
Since the phase space is fully noncommutative in terms of
these variables $x_0, p_0$, the usual formulation of quantum field
theory using a momentum representation will have to be examined.
We will call a phase space which satisfies
the relations \eq{fCR1}--\eq{fCR3} (not necessarily with $\k \propto
\th$) a deformed phase space. 
It will be very interesting to try to construct 
a field theory which represents the structure of this
deformed phase space.


\section{Discussions}

In this paper, we presented the quantization of the open string ending
on a D-brane in pp-wave background with a constant $B$-field. We found
that, due to the combined effects of the $B$-field and the mass
parameter, the noncommutative structure of the theory takes on a  new
form. 
Compared with the case of a noncommutative D-brane in a 
flat background, the novelty of the noncommutative algebra we
obtained in this paper is the noncommutativity
of the linear momentum:  (\ref{PP1}) at the level of the string
coordinates, and \eq{zCR3} at the level of the zero modes.
And moreover 
the boundary commutation relations 
depend on the lightcone momentum $p^+$ of the string.
In fact, the momentum is also noncommutative
for charged open strings, and similarly,
for open strings with their endpoints ending
on different D-branes with different background fields ${\cal F}$.
Since no state can be a simultaneous
eigenstate for all components of the momentum,
it is unclear whether one can use the usual Fourier
analysis to define the $\ast$-product as before.
(See \cite{Yin} for a proposal for the case of
open strings ending on D-branes with different backgrounds.)
Furthermore, in all of these cases, the linear momentum
is not a conserved quantity. Rather, the angular
momentum is conserved.
It would be very interesting if this leads to
a new class of noncommutative field theories and to investigate their
properties.

As in the neutral case, the novelty of the noncommutativity resides in
the zero modes sector. 
In the neutral string case, the effect of noncommutativity of the zero
modes in the operator formalism was  noticed in \cite{CZ}.  This led
directly to the construction of the general multiloop amplitudes for a
noncommutative open string using the Reggeon formalism \cite{CRS}.
In the charged string case, the noncommutativity of the zero modes 
played an important role in the determination of the 
charged string spectrum \cite{callan} and the partition function \cite{bachas}.
Moreover the 
charged string propagator has been constructed \cite{dn} 
which utilized essentially the noncommutativity of 
the zero modes. 
Aspects in the understanding of the charged string
interaction were recently made in \cite{CRS2}.
In our present case, we have determined some immediate physical 
effects of the commutation relations we obtained in this paper.
In terms of the oscillator algebra, \eq{a1}, \eq{a20} or 
\eq{a2}, it implies a splitting in the zero mode spectrum.
It also implies a more complicated mode number 
dependence of the nonzero mode spectrum.
Written in terms of the phase space variables, 
the zero modes noncommutativity takes the form \eq{zCR1}--\eq{zCR3} 
generically  depending
on the lightcone momentum of the string and is thus, in some sense, 
probe dependent. This is an interesting phenomena. 
A probe dependent phase space  may sound strange. 
However if geometry can be probe dependent (see for example \cite{greene} for
an introduction to various aspects of quantum geometry), it may not be
unnatural to expect an extension to that of a probe dependent phase
space. This is a
fundamental problem. We have not understood
its significance.
However it is an intriguing possibility. 
Although our results 
were derived in a background without gravity, it is
conceivable that quantum gravity
at the Planck scale may also involve a similar problem, and therefore
lessons that can be drawn from this construction should be helpful.

Naively, the commutation relations of the
deformed phase space
lead to the 
phase space uncertainty relations
\be \label{pur}
\Delta x^i \Delta p^j \sim |h^{ij}|,\quad
\Delta x^i \Delta x^j \sim |\th^{ij}|,\quad
\Delta p^i \Delta p^j \sim |\k^{ij}|.
\ee
Along the lines of \cite{ur},
this might be used to generalize
the spacetime uncertainty relations of Yoneya (see
\cite{yoneya} for a review and further references).
It would be
interesting to see how string dualities may put constraints
(e.g. string coupling dependence \cite{ur}) on the
forms of these phase space uncertainty relations.
We remark that \eq{pur} does not fall into
the classification of quantum
spacetime of \cite{DFR}. It
would be interesting to investigate what kinds of assumptions on the
quantum phase space could lead to the above uncertainty relations. 

Due to the  $p^+$-dependence  in some of our commutation relations, 
it is 
not clear whether one can apply the Mandelstam lightcone formalism. 
It is important to understand this issue.

Assuming we can overcome the problem discussed above and have 
gained sufficient control over string scattering in the GS
formulation, it will also be very 
interesting to understand the results obtained in
this paper, particularly the momentum noncommutativity,
in the approach of Schomerus \cite{Schom}.
This will be useful for the understanding of  the properties of 
the D-brane low energy theory.

\acknowledgments{
CSC thanks Simon Ross for helpful discussions and Rodolfo Russo for useful
comments. 
The work of CSC was supported in part by the Nuffield Foundation,
Grant Number NUF-NAL/00445/G. 
The work of PMH is supported in part by
the National Science Council,
the Center for Theoretical Physics
at National Taiwan University,
the National Center for Theoretical Sciences,
and the CosPA project of the Ministry of Education,
Taiwan, R.O.C.
}

\ed
